\documentclass[fleqn,usenatbib,letters]{mnras}

\usepackage{newtxtext,newtxmath}
\usepackage[T1]{fontenc}

\DeclareRobustCommand{\VAN}[3]{#2}
\let\VANthebibliography\thebibliography
\def\thebibliography{\DeclareRobustCommand{\VAN}[3]{##3}\VANthebibliography}

\usepackage{graphicx}	% Including figure files
\usepackage{amsmath}	% Advanced maths commands
\usepackage{amssymb}	% Extra maths symbols
\usepackage{gensymb}

\usepackage{symbols}
\title[H{\sc{i}} scale heights in simulated $\Ls$ galaxies]{Realistic \hi scale heights of Milky Way-mass galaxies in the FIREbox cosmological volume}

\author[Jindra Gensior et al.]{
\parbox{\textwidth}{Jindra Gensior,$^{1}$\thanks{E-mail: jindra.gensior@uzh.ch}
Robert Feldmann,$^{1}$
Lucio Mayer,$^{1}$
Andrew Wetzel,$^{2}$
Philip F. Hopkins,$^{3}$
and Claude-Andr{\'e} Faucher-Gigu{\`e}re$^{4}$
}\\
$^{1}$Institute for Computational Science, University of Z{\"u}rich, Winterthurerstrasse 190, 8057 Z{\"u}rich, Switzerland\\
$^{2}$Department of Physics \& Astronomy, University of California, Davis, CA 95616, USA\\
$^{3}$California Institute of Technology, TAPIR, Mailcode 350-17, Pasadena, CA 91125, USA\\
$^{4}$CIERA and Department of Physics and Astronomy, Northwestern University, 1800 Sherman Ave, Evanston, IL 60201, USA\\
}

\date{Accepted 2022 October 26. Received 2022 October 7; in original form 2022 July 7}

\pubyear{2022}

\begin{document}
\label{firstpage}
\pagerange{\pageref{firstpage}--\pageref{lastpage}}
\maketitle

% Abstract of the paper
\begin{abstract}
Accurately reproducing the thin cold gas discs observed in nearby spiral galaxies has been a long standing issue in cosmological simulations. Here, we present measurements of the radially resolved \hi scale height in 22 non-interacting Milky Way-mass galaxies from the \fc box cosmological volume. We measure the \hi scale heights using five different approaches commonly used in the literature: fitting the vertical volume density distribution with a Gaussian, the distance between maximum and half-maximum of the vertical volume density distribution, a semi-empirical description using the velocity dispersion and the galactic gravitational potential, the analytic assumption of hydrostatic equilibrium, and the distance from the midplane which encloses $\gtrsim$60~per~cent of the \hi mass. We find median \hi scale heights, measured using the vertical volume distribution, that range from $\sim$100 pc in the galactic centres to $\sim$800 pc in the outskirts and are in excellent agreement with recent observational results. We speculate that the presence of a realistic multiphase interstellar medium, including cold gas, and realistic stellar feedback are the drivers behind the realistic \hi scale heights. 
\end{abstract}

% Select between one and six entries from the list of approved keywords.
% Don't make up new ones.
%%%%%%%%%%%%%%%%%%%%%%%%%%%%%%%%%%%%%%%%%%%%%%%%%%

%%%%%%%%%%%%%%%%% BODY OF PAPER %%%%%%%%%%%%%%%%%%

\section{Introduction}\label{s:intro}
\vspace{-0.15cm}
Producing realistic disc galaxies in cosmological scale simulations has been a long-standing problem for numerical astrophysics \citep[e.g.][and the references therein]{Naab2017}. Only in the past decade have cosmological zoom-in simulations and volumes been able to reproduce the rotation curves and thin stellar discs seen in late-type spiral galaxies \citep[e.g.][]{Guedes2011,Stinson2013,Aumer2013,Hopkins2014,Agertz2015,Grand2017,Ma2017,Garrison-Kimmel2018,Pillepich2019,Buck2020}. This success if often attributed to the (stellar) feedback models used, as the thickness of the stellar disc is sensitive to the feedback strength \citep[e.g.][]{Sokolowska2017}.

Resolving the gas scale height has been identified as another crucial part of reproducing realistic Milky Way-like, $\Ls$ galaxies \citep{Hopkins2018}. Observations indicate that the \hi discs of galaxies are very thin, with scale heights of $\sim$100 pc in the centre, flaring to $\lesssim$1 kpc at large galactic radii \citep{Yim2011,Yim2014,Bacchini2019a,Patra2020,Randriamampandry2021}. \citet{Randriamampandry2021} found average scale heights within the optical radius of the spiral galaxies studied of $\sim$0.35 $\pm$ 0.16 kpc for the \bd \citep{Wang2013} galaxies and $\sim$0.68 $\pm$ 0.58 kpc for The \hi Nearby Galaxy Survey \citep[THINGS;][]{Walter2008}. 

By contrast, recent cosmological (zoom) simulations which investigated the \hi content of $\Ls$ galaxies found the \hi discs to be much thicker. \hi scale heights of the Auriga \citep{Grand2017}  galaxies within the optical radius range from $\sim$1${-} 10~\kpc$ \citep{Marinacci2017}. Similarly, the EAGLE \citep[][]{Schaye2015} galaxies have \hi discs with scale heights > 1.5 kpc, which \citet{Bahe2016} interpreted as consequence of too strong feedback. 
The scale height or thickness of gas discs is relevant for determining their susceptibility to gravitational instability \citep[e.g.][]{Romeo2011}. Whether a (super)bubble driven by supernovae (SNe) can break out of the galactic disc and pollute the circumgalactic medium (CGM) with metals, or whether it stalls and deposits all ejecta in the gas disc of the galaxy also depends on the scale height of the cold gas \citep[e.g.][]{Silich2001,Orr2021,Orr2022}. Thus the scale height has direct implications for the star formation and feedback cycle of galaxies, as well as the metal pollution of the CGM. In turn, feedback will affect the \hi structure  \citep[e.g.][]{Bahe2016}. Therefore, the \hi scale height can be a sensitive test for sub-grid physics, as it is unclear a priori whether well validated sub-grid models like \fc -2 will be able to reproduce observed scale heights. Accurately modelling (spiral) galaxies and their \hi gas content is also highly relevant with respect to the upcoming Square Kilometer Array \citep[SKA;][]{Dewdney2009} observatory and its precursors. Both to make higher fidelity predictions, and to be able to better interpret observational results. 

Encouragingly, thin cold gas discs have been found in several $\Ls$ galaxies simulated with the \fc -2 model \citep{Sanderson2020,Gurvich2020,Trapp2022}, but no systematic exploration with a focus on the HI has been conducted yet. The \fcb cosmological volume \citep{Feldmann2022}, with its high spatial resolution ($\sim$20 pc), and multiphase ISM, that yields 20{--}30 $\Ls$ galaxies, is ideally suited to systematically analyse the \hi discs of $\Ls$ galaxies. In this letter, we focus on their scale heights in particular, measuring them with the variety of different methods used in the literature.

\vspace{-0.75cm}

\section{FIREbox}\label{ss:methFB}
\vspace{-0.15cm}
In this letter we study galaxies drawn from the (22.1 Mpc)$^3$ \fc box cosmological volume simulation \citep{Feldmann2022} that is part of the Feedback In Realistic Environments (\fc) project\footnote{\url{https://FIRE.northwestern.edu/}}.The simulation was run with the meshless-finite-mass code {\sc{Gizmo}}\footnote{\url{http://www.tapir.caltech.edu/~phopkins/Site/GIZMO.html}} \citep{Hopkins2015} and the \fc -2 \citep{Hopkins2018} sub-grid physics. Specifically, these simulations use the \citet{Hopkins2014} cooling and heating rates valid for temperatures ranging from $10 {-} 10^9~\K$, including heating and photoionisation from a \citet{FaucherGiguere2009} UV Background, naturally resulting in a multiphase interstellar medium (ISM) that includes a cold gas phase. Stars form with a local efficiency of 100~per~cent from gas that is molecular, self-gravitating, Jeans unstable and above a density threshold of $n\geq300~\ccm$. A variety of stellar feedback channels are included, namely SN Type II and Ia, stellar winds from massive OB and evolved AGB stars, as well as photoionization, photoelectric heating and radiation pressure. 

\fcb has a mass resolution of $m_{\rm b}=6.3\times10^4~\Msun$ ($m_{\rm DM}=3.3\times10^5~\Msun$), fixed gravitational softening for the stars (12 pc) and dark matter (80 pc), and adaptive gravitational softening for the gas down to a softening length of 1.5 pc (all in proper units). Galaxies were identified at $z=0$ using the AMIGA halo finder \citep{Gill2004,Knollmann2009}. We restrict our analysis to centrals in the virial mass range $11.85 < \log_{10}(\Mvir/\Msun) < 12.48$, such that the mean of the resultant 27 galaxies matches the Milky Way $\Mvir = 1.3\pm0.1\times 10^{12}$ from \citet{Bland-Hawthorn2016}. To better match the observational samples, we visually inspect the galaxies and exclude those that are interacting or undergoing a merger. The stellar masses of our final sample of 22 star-forming, late-type galaxies range from $3.5\times10^{10}$ to $1.6\times10^{11}~\Msun$, in good agreement with the mass range of the \bd galaxies and the higher mass THINGS galaxies studied by \citet{Patra2020}, while \citet{Bacchini2019a} also include some lower mass THINGS galaxies.

To obtain the \hi mass fraction of each gas particle, we first calculate the $\Ht$ fraction of the neutral gas using Equation 1 of \citet{Krumholz2011}, from the metallicity and dust optical depth (estimated from the metallicity and density) of the particle. We then subtract the molecular fraction, as well as contributions from metals and Helium, from the neutral gas fraction of each particle.
\vspace{-0.3cm}

\vspace{-0.35cm}
\section{Measuring the HI scale height}\label{ss:methhHI}
\vspace{-0.15cm}
The scale height of real galaxies can only be measured directly for edge-on galaxies, where the vertical gas distribution can be fit with a Gaussian \citep[e.g.][]{Yim2011}.
Below, we briefly discuss the different ways the \hi scale height can be measured for galaxies at lower inclinations. 

Assuming that the gas disc is in hydrostatic equilibrium, with turbulent pressure\footnote{\citet{Bacchini2019a} explicitly assume the gas to be vertically isothermal. However, it has also been shown in numerical simulations with effective feedback that the turbulent pressure dominates over the thermal pressure \citep[][]{Benincasa2016}.} balancing the effect of gravity, it is possible to derive an \emph{analytic estimate} of the \hi scale height \citep[see Appendix A of][and references therein]{Bacchini2019a}.
\begin{equation}\label{eq:analyticalHIs}
    \centering 
    \hhi(R) \approx \sigma_{\rm HI}(R)/\sqrt{4\pi G\times\left(\rho(R,0) - \frac{1}{2\pi G}\frac{v_{\rm c}(R)}{R}\frac{\partial v_{\rm c}(R)}{\partial R}\right)},
\end{equation}where $\hhi$ is the scale height of the \hi disc, $\sigma_{\rm HI}$ = $\sqrt{\langle(v - \langle v\rangle)^2\rangle/3}$ is the \hi mass-weighted line-of-sight velocity dispersion in the direction perpendicular to the disc plane, $v_{\rm c}$ is the circular velocity and $\rho(R,0)$ is the mean mass-weighted density at the galaxy mid-plane, all as a function of galactocentric radius $R$. We calculate all quantities as running mean in overlapping radial bins of 1 kpc width.

The analytic approximation neglects the contribution of the gas self-gravity to the gravitational potential. To include gas self-gravity, and thus determine $\hhi$ more accurately, one needs to solve for the volume density numerically. It is done by solving the equation of the hydrostatic equilibrium condition under the constraints of the rotation curve and velocity dispersion, iterating until the so derived estimates converge. This method has been used in two subtly-different ways in the literature.

The first approach assumes that the vertical volume density profile has a \emph{Gaussian} shape that depends on the \hi scale height, i.e. $\rho(z) \propto \exp(-z^2/(2\hhi^2))$. It was used by e.g. \citet{Bacchini2019a} to measure the \hi scale heights of 12 THINGS galaxies. To mimic this method, we fit the \hi mass-weighted vertical volume density profile of gas with $z\leq2~\kpc$ in radial annuli of 1 kpc width, with a Gaussian function to measure $\hhi$.

With the \emph{half-width half maximum} (HWHM) approach, no functional form is assumed for the \hi volume density. The \hi scale height is defined as the distance between the midplane (maximum volume density) and one of the points where the \hi volume density reaches its half-maximum. It was used by e.g. \citet{Patra2020} to measure $\hhi$ in 9 THINGS galaxies. We determine this distance for the \hi mass-weighted vertical volume density profile of gas with $z\leq2~\kpc$ in radial annuli of 1 kpc width.

Another way to estimate $\hhi$ using a \emph{semi-empirical formula} was originally derived by \citet{Wilson2019}, assuming hydrostatic equilibrium but taking into account observational results regarding the gravitational potentials of galaxies. We follow the parametrisation of \citet{Randriamampandry2021}, who used this equation to measure $\hhi$ for the \bd galaxies. The \hi scale height can be expressed as:
\begin{equation}\label{eq:Wilsonf}
    \centering
    h_{\rm HI}(R) = \frac{\sigma_{\rm HI}^{2}(R)}{\pi G \Sigma_{\rm total}} \times \left({1 + \sigma^{2}_{\rm HI, 0}}/{\left[2v_{\rm max}^{2} \times \left(\frac{M_{\rm dyn}}{M_{\rm HI} + \Mstar}\right)^2\right]}\right)^{-1}\hspace{-0.9em},
\end{equation}where $\sigma_{\rm HI, 0}$ is the central \hi velocity dispersion, $\Sigma_{\rm total}$ is the total surface density within the gas layer, $v_{\rm max}$ is the peak of the galaxy's rotation curve, $M_{\rm dyn}$ the total (dynamical) mass, $M_{\rm HI}$ the total mass of \hi and $\Mstar$ the total stellar mass. We calculate all radially dependent quantities in radial annuli of 1 kpc width and the masses within $0.1\Rvir$ of the galaxy, in analogy to the optical radius used by \citet{Randriamampandry2021}.

Finally, \citet{Marinacci2017} measure the \hi scale height as the distance from the midplane where the \hi \emph{mass enclosed} within a cylinder spanning $\pm30~\kpc$ from the midplane of the galaxy equals 75~per~cent of the \hi mass. We calculate the distance from the midplane that encloses 63~per~cent (one e-folding) of the \hi mass within $30~\kpc$ above and below the midplane in radial annuli of 1 kpc width, to determine how this method of measuring $\hhi$ compares to the other approaches used in the (observational) literature. To better compare to the Auriga scale height measurements, we also measure a single scale height in a solid cylinder with radius $0.1\Rvir$.

\vspace{-0.5cm}
\section{Properties of FIREbox galaxies}\label{s:galprops}
\vspace{-0.15cm}
\begin{figure}
    \centering
    \includegraphics[trim=1.5mm 3mm 1.5mm 2.6mm, clip=true,width=\linewidth]{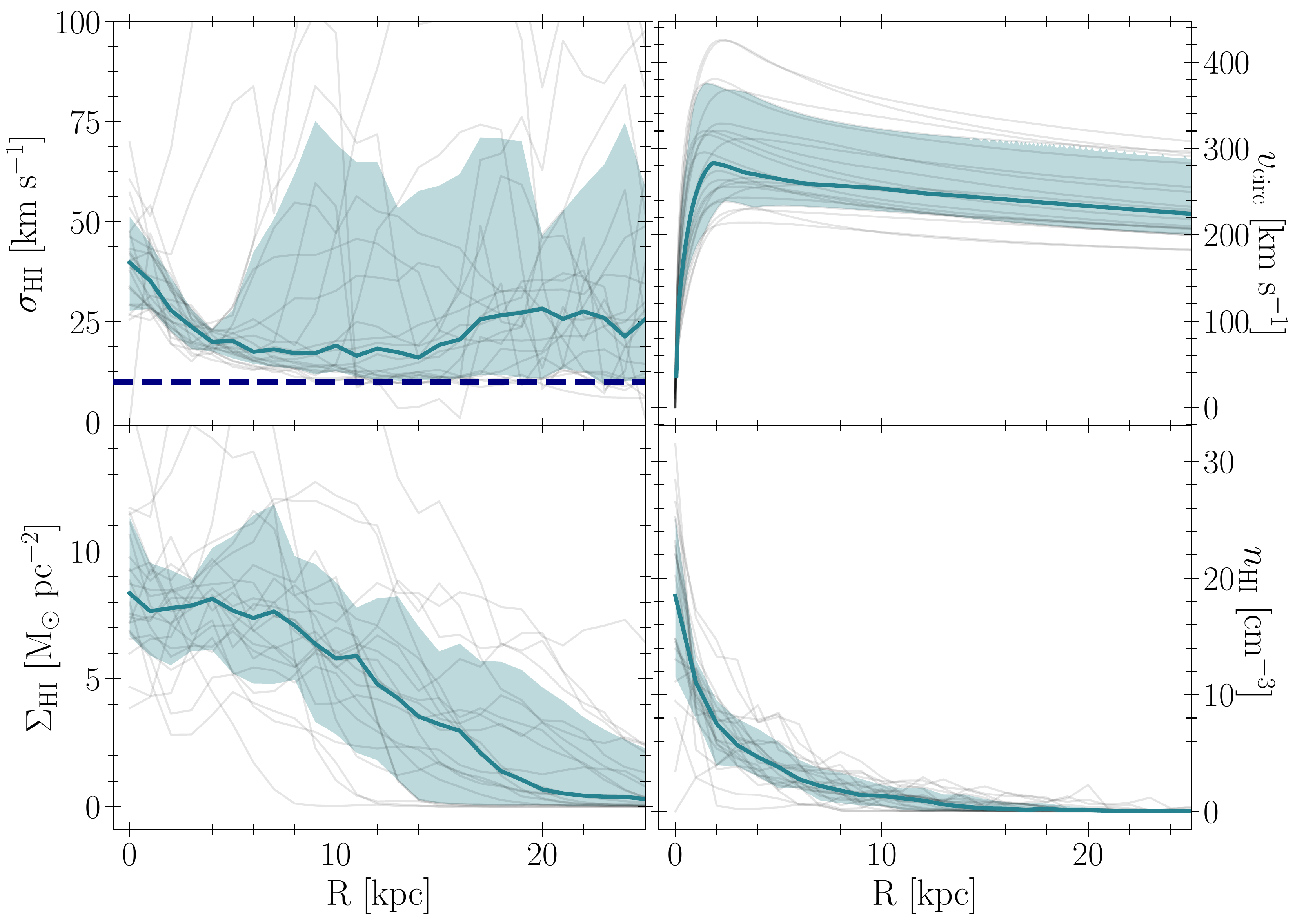}
    \vspace{-0.5cm}
    \caption{Properties of the 22 \fc box galaxies considered in this letter: \hi velocity dispersion (top left), circular velocity (top right), \hi mass surface density (bottom left) and \hi volume density (bottom right) as a function of radius. Thin grey lines show the radial profiles of the individual galaxies. The blue line indicates the sample median, and the blue shaded region the 16th-to-84th percentile. The dashed navy line in the top left plot indicates a velocity dispersion of $10~\kms$.\vspace{-0.4cm}}
    \label{fig:gal_properties}
\end{figure}
We show the radial profiles of several galaxy properties (\hi velocity dispersion, \hi surface density, \hi mass-weighted volume density at the midplane and rotation velocity curves), relevant for calculating the \hi scale height either analytically (Equation~\ref{eq:analyticalHIs}) or using the \citet{Wilson2019} semi-empirical equation (\ref{eq:Wilsonf}), in Figure~\ref{fig:gal_properties}. The median $\sigma_{\rm HI}$ radial profile peaks at the centre of the galaxy with 40 $\kms$, declining with increasing galactocentric radius to a value of $\sim17~\kms$ for $\rm R<15$ kpc, where it increases again to $\sim27~\kms$. The median consistently lies above the often assumed canonical value of $\sigma_{\rm HI}\sim10~\kms$ \citep[e.g.][]{Tamburro2009}, also seen in the \bd galaxies \citep{Randriamampandry2021}. However the velocity dispersion for the \fc box galaxies is calculated for all \hi gas, including the warm component, for which \citet{Ianjamasimanana2012} found an average velocity dispersion of $16.8~\kms$ in good agreement with the median $\sigma_{\rm HI}$ at intermediate radii. The rise of the median velocity dispersion in the central 5 kpc is indicative of shear-driven turbulence induced by a bulge (see e.g. Figure 9 in \citealt{Gensior2020}). The median rotation curve in the top right panel shows the characteristic bump expected from a spheroidal component around $R = 2~\kpc$, supporting this interpretation.
The velocity dispersion profiles of the THINGS galaxies with bulges studied in \citet{Bacchini2019a} and \citet{Patra2020} qualitatively match the radial profiles in Figure~\ref{fig:gal_properties}, although the central velocity dispersions of the THINGS galaxies reach $\sim30~\kms$ at most. 
The \hi surface density is calculated as the \hi mass enclosed in overlapping radial annuli of width 1 kpc, divided by the area of the annulus. The median $\Sigma_{\rm HI}$ stays approximately constant at $\sim$7.5 $\Msun\pc^{-2}$ out to a galactocentric radius of 8 kpc, before declining to a density $<1\Msun\pc^{-2}$ at a median \hi disc radius $R_{\rm HI}\approx$ $20^{+9}_{-7}$ kpc, in good agreement with observed $\Sigma_{\rm HI}$ profiles \citep[e.g.][]{Leroy2008,Wang2014,Randriamampandry2021}. 
The median \hi mass-weighted volume density at the midplane peaks at 18 $\ccm$ in the galactic centre and declines steeply with increasing galactocentric radius, falling below 1 $\ccm$ at a radius of 13 kpc.

\vspace{-0.5cm}
\section{\hi scale height of FIREbox galaxies}\label{s:hHI}
\vspace{-0.15cm}
\begin{figure*}
    \centering
    \includegraphics[trim=1.5mm 3mm 1.5mm 5.5mm, clip=true,width=\linewidth]{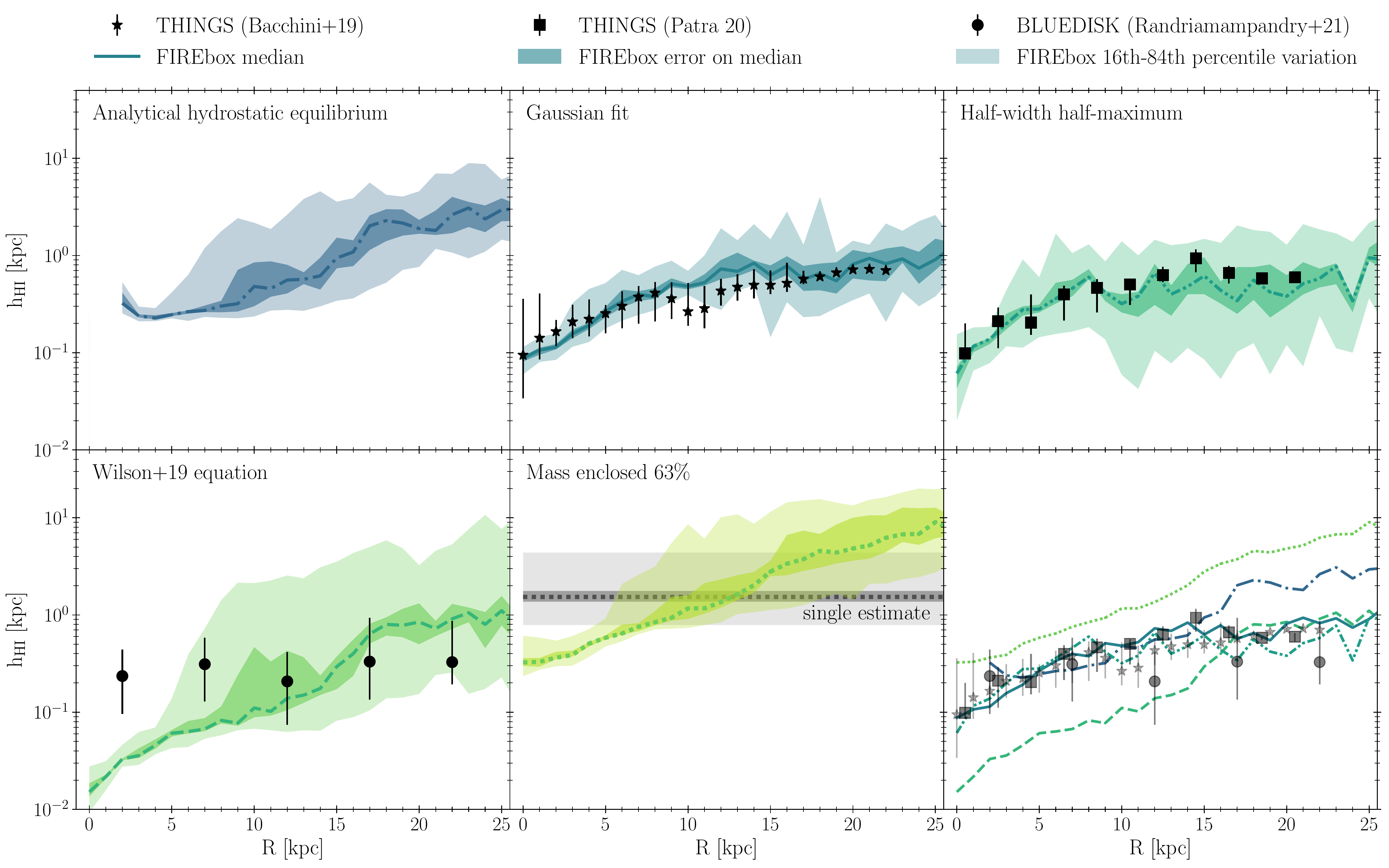}
    \caption{Radial profiles of the \hi scale heights of the \fc box $\Ls$ galaxies estimated from Equation~\ref{eq:analyticalHIs}, \ref{eq:Wilsonf}, a Gaussian fit to the vertical volume density profile, distance from the galaxy mid-plane that encloses 63~per~cent of the \hi mass (including a single measurement in a cylinder of R = $0.1\Rvir$) and the vertical distance between the maximum and half of the maximum volume density, from top left to bottom left respectively, with observations from \citet{Randriamampandry2021}, \citet{Bacchini2019a}, and \citet{Patra2020} over-plotted, where the same method of estimating the \hi scale height is used. Coloured lines indicate the median, the dark shaded regions the error on the median determined via bootstrapping and the light shaded regions the 16th-to-84th percentile variation of the data. The bottom right panel combines the median measurements of observations and simulations presented in the other five panels. The \hi scale heights measured from the vertical volume density distribution are in excellent agreement with observational measurements.\vspace{-0.5cm}}
    \label{fig:hHI}
\end{figure*}

Figure~\ref{fig:hHI} shows the \hi scale height of the 22 $\Ls$ \fc box galaxies measured using the different approaches outlined in Section~\ref{ss:methhHI}. The top three and bottom two leftmost panels show the measurements for individual galaxies, with median, errror on the median determined via bootstrapping, and 16th-to-84th percentile variation for the sample indicated by the coloured line and dark and light shaded regions respectively, compared with the observational result using the same method where applicable. The bottom right panel shows a compilation of the median measurements from the different methods, simulations and observations. 

All median \hi scale height profile obtained from the simulated galaxies show disc flaring, i.e. follow the trend of increasing $\hhi$ with increasing galactocentric radius. However, the different methods vary by more than an order of magnitude in their estimate of the central $\hhi$, the amount of disc flaring predicted, as well as the radial gradient of the profile. 

Using the vertical profile of the \hi mass-weighted volume density to determine the $\hhi$, either by fitting the gas distribution with a Gaussian, or calculating the HWHM, yields results in excellent agreement with measurements from the THINGS galaxies that use the same methodology \citep{Bacchini2019a,Patra2020}. The HWHM method finds a lower central scale height compared to the Gaussian fit (60 and 85 pc respectively), and predicts a steeper increase to a scale height of 500 pc at $\rm R=8~\kpc$, before oscillating and gradually increasing to 900 pc at $\rm R=25~\kpc$. The \hi scale height obtained from the Gaussian fit increases more shallowly with R, reaching 800 pc at $\rm R=14~\kpc$, staying nearly constant at larger radii (reaching 1 kpc at $\rm R=25~\kpc$). 
The analytic determination of $\hhi$ agrees well with the volume density estimates and the observations for $\rm R \leq 15~\kpc$, but predicts too much flaring at larger galactocentric radii, likely driven by the increase of $\sigma_{\rm HI}$ beyond this radius.

Determining the \hi scale height from Equation~\ref{eq:Wilsonf} yields a median central scale height of 13 pc, increasing to 100 pc at a galactocentric radius of 12 kpc, and flaring further until reaching a scale height of $\approx0.9~\kpc$ at $\rm R>17.5~\kpc$. The median of the simulated galaxies lies significantly below the median \citet{Randriamampandry2021} \bd measurement, that only shows a very modest increase from 200 to 350 pc, for $R<12~\kpc$. Due to the large flaring predicted, the median of the simulated galaxies lies within the 16th-to-84th percentile of \bd at larger radii. The large discrepancies at small galactocentric radii are a consequence of the strong dependence of $\hhi$ on $\sigma_{\rm HI, 0}$ in Equation~\ref{eq:Wilsonf}, as the central velocity dispersions of the \fc box $\Ls$ galaxies far exceed the $\sim10~\kms$ of the \bd galaxies\footnote{However, the \bd observations have coarse resolution and lack $\sigma$ measurements in their centres, making it difficult to determine how much of the discrepancy between the \bd observations and simulations is due to (potential) morphological differences or strong \fc -2 feedback potentially further enhancing the velocity dispersion.}. At large galactocentric radii the strong flaring is driven is driven by the increase in the median $\sigma_{\rm HI}$ from 17 to 27 $\kms$ (see the top left panel of Figure~\ref{fig:gal_properties}).  
Defining the scale height as the distance from the midplane that encloses 63~per~cent of the total \hi mass in a hollow cylinder of height 60 kpc systematically overpredicts $\hhi$ compared to all other methods of obtaining a scale height, with a median scale height of 350 pc at the galactic centre, increasing to 10 kpc by a galactocentric radius of 25 kpc. The median $\hhi$ measured in a solid cylinder of radius $0.1\Rvir$ is 1.53 kpc, highlighting that this single measurement also overestimates the \hi scale height. The total \hi mass enclosed depends on the height of the cylinder, thus using a smaller distance from the midplane would yield lower scale height estimates.

\vspace{-0.6cm}
\section{ Why does FIREbox produce realistic HI scale heights?}\label{s:tdc}
\vspace{-0.15cm}
We can speculate what makes \fc box so successful at producing the thin \hi discs shown in Figure~\ref{fig:hHI} based on possible explanations put forward in the literature. In some cases \citep[e.g.][]{Guedes2011,Pillepich2019} high resolution has been crucial to reproducing thin (stellar) discs. \fcb has a mass resolution comparable to that of other state-of-the-art cosmological (zoom) simulations like IllustrisTNG50 \citep[][]{Pillepich2019} and Auriga \citep[][]{Grand2017}, but allows for higher spatial resolution due to its smaller minimum gravitational softening. However, as we show in the online supplementary material, using the vertical volume density profile to determine the \hi scale height yields good agreement for simulations differing by nearly two orders of magnitude in spatial resolution (minimum gas softening ranging from 1.5 to 100 pc). This suggests that the other modelling choices are more relevant for producing realistic galaxy discs. However, the minimum gravitational softening likely needs to be lower than the scale height, to reproduce thin scale heights. This is consistent with the \fcb results, as the lowest resolution re-run has a minimum gravitational force softening below the smallest measured $\hhi$ of $\sim$100 pc.\\ 
\indent \citet{Trayford2017} suggest that using an effective equation of state instead of directly modelling the multiphase ISM leads to too thick stellar discs in the EAGLE galaxies. \citet{Pillepich2019} also report a 0.5 dex variation in scale height when adjusting the parameters of their equation of state model. The \fc -2 model includes cooling down to temperatures of 10 K, thus we can self-consistently model a multiphase ISM including cold gas, plausibly a major factor in producing realistic gas scale heights. \\
\indent Many authors have argued that the inclusion of more sophisticated feedback models such as superbubble feedback \citep[e.g.][]{Keller2015} or early stellar feedback are key to produce realistic $\Ls$ galaxies (e.g. \citealt{Stinson2013,Aumer2013,Hopkins2014}). 
This suggests that the early stellar feedback implemented in \fc -2 could be crucial to producing thin gas discs, by pre-processing the ISM gas, thus enabling the effective driving of localised SNe outflows even at coarse mass resolution (see Figure~A1 in the online supplementary material), as opposed to SNe depositing more of their energy and momentum within the ISM, increasing the local pressure and thereby increasing the overall scale height. Apart from dedicated new simulations, it might be informative to see whether other simulations with a cold ISM but no early stellar feedback (e.g. EMP-\emph{Pathfinder}; \citealt{ReinaCampos2022}) and those with early stellar feedback \citep[e.g. NIHAO;][]{Buck2020} also show thin \hi gas discs. 

\vspace{-0.5cm}
\section{Concluding remarks}
\vspace{-0.15cm}
We measure the \hi scale height of 22 Milky Way-mass galaxies from the \fc box pathfinder volume, using five different methods. We find that the two methods using the vertical volume density distribution (Gaussian fit and HWHM) are in excellent agreement with each other and the observations. We speculate that the combination of detailed modelling of the cold ISM and early stellar feedback are responsible for these realistically thin \hi gas discs, a trend that is independent of resolution.

The analytic hydrostatic equilibrium assumption works well for the central $\sim15~\kpc$, but overpredicts the scale height compared to observations and measurements from the vertical volume density distribution at large radii, due to the rising velocity dispersion. The \citet{Wilson2019} equation underpredicts the scale height in the central region of the galaxies, due to the high central velocity dispersion of the \fcb galaxies, but is in good agreement with the volume density estimates at large radii. Using the \hi mass enclosed as an estimate consistently overpredicts the gas scale height at all radii, although the general trend of disc flaring at large radii is present as well.

Turbulent theory predicts that the \hi scale height is a transition scale, where turbulence changes from three-dimensional turbulence on small scales to two-dimensional turbulence on larger scales \citep[e.g.][]{Agertz2015}. This could cause a break in the \hi power spectra of galaxies \citep[e.g.][]{Dutta2008}. Since \fc box produces galaxies with realistic \hi scale heights, it is well suited to study \hi turbulence and its drivers and other processes that depend on the scale height. Futhermore, this makes \fc box a great resource for more in-depth studies of the \hi content of galaxies in general, and for making predictions for upcoming observatories such as the SKA.

\vspace{-0.5cm}
\section*{Acknowledgements}
\vspace{-0.15cm}
We thank an anonymous referee for their helpful and constructive report. JG, RF and LM gratefully acknowledge financial support from the Swiss National Science Foundation (grant no CRSII5\_193826). RF acknowledges financial support from the Swiss National Science Foundation (grant no PP00P2\_194814 and 200021\_188552). AW received support from: NSF via CAREER award AST-2045928 and grant AST-2107772; NASA ATP grant 80NSSC20K0513; HST grants AR-15809, GO-15902, GO-16273 from STScI. CAFG was supported by NSF through grants AST-1715216, AST-2108230,  and CAREER award AST-1652522; by NASA through grants 17-ATP17-006 7 and 21-ATP21-0036; by STScI through grants HST-AR-16124.001-A and HST-GO-16730.016-A; by CXO through grant TM2-23005X; and by the Research Corporation for Science Advancement through a Cottrell Scholar Award. We acknowledge PRACE for awarding us access to MareNostrum at the Barcelona Supercomputing Center (BSC), Spain. This work was supported in part by a grant from the Swiss National Supercomputing Centre (CSCS) under project IDs s697 and s698. We acknowledge access to Piz Daint at the Swiss National Supercomputing Centre, Switzerland under the University of Zurich's share with the project ID uzh18. This work made use of infrastructure services provided by S3IT (www.s3it.uzh.ch), the Service and Support for Science IT team at the University of Zurich. 

%%%%%%%%%%%%%%%%%%%%%%%%%%%%%%%%%%%%%%%%%%%%%%%%%%
\vspace{-0.65cm}
\section*{Data Availability}
\vspace{-0.15cm}
The data underlying this article will be shared on reasonable request to the corresponding author.

%%%%%%%%%%%%%%%%%%%% REFERENCES %%%%%%%%%%%%%%%%%%

% The best way to enter references is to use BibTeX:
\vspace{-0.6cm}
\bibliographystyle{mnras}
\bibliography{bibliography} % if your bibtex file is called example.bib

% Alternatively you could enter them by hand, like this:
% This method is tedious and prone to error if you have lots of references
%\begin{thebibliography}{99}
%\bibitem[\protect\citeauthoryear{Author}{2012}]{Author2012}
%Author A.~N., 2013, Journal of Improbable Astronomy, 1, 1
%\bibitem[\protect\citeauthoryear{Others}{2013}]{Others2013}
%Others S., 2012, Journal of Interesting Stuff, 17, 198
%\end{thebibliography}

%%%%%%%%%%%%%%%%%%%%%%%%%%%%%%%%%%%%%%%%%%%%%%%%%%

%%%%%%%%%%%%%%%%% APPENDICES %%%%%%%%%%%%%%%%%%%%%

\appendix
\section{Resolution Test}\label{A:res}
We study the effect of spatial and mass resolution on the results presented in this letter by applying the analysis described in Section~3 to two lower resolution re-runs of \fcb, listed in Table~\ref{tab:FBres}. Following the selection criteria discussed in Section~2, this results in 23 $\Ls$ galaxies for the intermediate resolution volume and 25 $\Ls$ galaxies for the low resolution volume. Additionally, we use 7 higher resolution Milky-Way cosmological zoom-in simulations from the \fc -2 suite (m12s), to gain an intuition about convergence at and above the \fcb resolution. 
\begin{table}
    \caption{Properties of the simulations}
    \centering
    \begin{tabular}{c|c|c|c|c|c|c}
        \hline
        Name & \# cells & $M_{\rm b}$ & $\epsilon^{\rm min}_{\rm gas}$ & $d_{\rm SF}$ & $\epsilon_{\ast}$ & $n_{\rm thresh}$\\
        & & ($\Msun$) & (pc) & (pc)& (pc) & ($\ccm$)\\
        \hline
        \fcb & $1024^3$ & $6\times10^4$ & 1.5 & 20 & 12 & 300\\
        FB512 & $512^3$ & $5\times10^{5}$ & 4 & 59 & 32 & 100\\
        FB256 & $256^3$ & $4\times10^{6}$  & 16 & 253 & 128 & 10\\
        \hline
        m12b & - & $7\times10^3$ & 0.38 & 7 & 4 & 1000 \\
        m12c & - & $7\times10^3$ & 0.38 & 7 & 4 & 1000 \\
        m12f & - & $7\times10^3$ & 0.38 & 7 & 4 & 1000 \\
        m12i & - & $7\times10^3$ & 0.38 & 7 & 4 & 1000 \\
        m12m & - & $7\times10^3$ & 0.38 & 7 & 4 & 1000 \\
        m12r & - & $7\times10^3$ & 0.38 & 7 & 4 & 1000 \\
        m12w & - & $7\times10^3$ & 0.38 & 7 & 4 & 1000 \\       
        \hline
        \multicolumn{7}{p{\linewidth}}{\footnotesize{\textit{Notes:} Column 1 lists the name of the simulation (volume), where the lower resolution \fcb re-runs are referred to by the abbreviation FB, followed by the number of mesh elements per side. Column 2 lists the number of cells, column 3 lists the baryonic mass resolution, column 4 lists the Plummer-equivalent minimum gas gravitational softening, column 5 lists the gas gravitational softening at the density threshold for star formation, column 6 lists the stellar gravitational softening and column 7 lists the density threshold for star formation.}} 
    \end{tabular}
    \label{tab:FBres}
\end{table}

\begin{figure*}
    \centering
    \includegraphics[width=\linewidth]{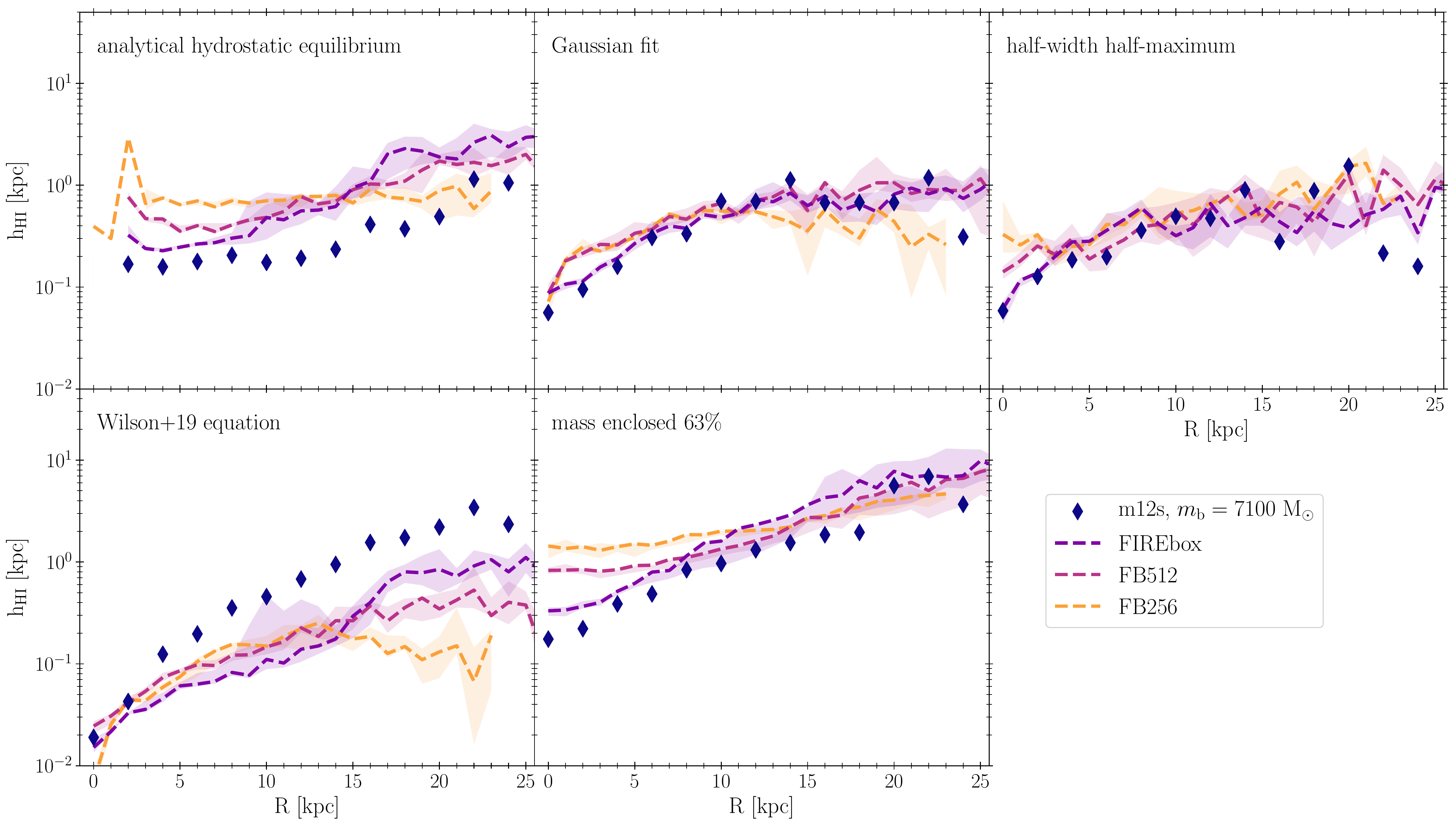}
    \caption{Comparing the effect of different spatial resolutions on the measured \hi scale heights of the \fc box $\Ls$ galaxies. As in Figure~2, from left to right, top to bottom, scale heights are calculated assuming hydrostatic equilibrium, from a Gaussian fit to the vertical volume density profile, the distance between the maximum and half-maximum volume density, Equation~2 and as distance from the mid-plane enclosing 63~per~cent of the total \hi mass. Coloured, dashed lines show the median of the samples: purple for the fiducial resolution \fc box, magenta for the intermediate resolution \fc box (FB512) and orange for the low resolution \fc box (FB256). Shaded regions indicate the error on the median determined via bootstrapping. Dark blue data points show the median scale height of the 7 higher resolution cosmological-zoom \fc -2 $\Ls$ galaxies (m12s). Especially at intermediate radii, the simulations show excellent convergence.}
    \label{fig:hHI_res}
\end{figure*}

We compare the different \hi scale height measurements for $\Ls$ galaxies at different resolutions in Figure~\ref{fig:hHI_res}. Apart from the \hi scale height obtained from Equation~2, the scale height tends to be overestimated in the inner 3{--}5 kpc at lower resolutions, by at most factor of two at the galactic centre in the lowest resolution \fc box (FB256). Beyond that the measurements converge, showing excellent agreement at intermediate galactocentric radii, although the low resolution \fc box galaxies tend to underpredict disc flaring at large galactocentric radii. The higher resolution simulations follow the resolution trend seen in the \fc boxes. Differences between the m12s with \fcb are most pronounced when using Equations~1 and 2 to estimate the scale height, a consequence of the differences in gravitational potential, velocity dispersion and surface density radial profiles of the m12s (potentially exacerbated by the smaller sample size). Scale heights measured from the vertical volume density distribution (Gaussian fit and half-width half-maximum), are well converged for \fcb and the m12s.  
We conclude that our results do not have a strong resolution dependence.  

%%%%%%%%%%%%%%%%%%%%%%%%%%%%%%%%%%%%%%%%%%%%%%%%%%

% Don't change these lines
\bsp	% typesetting comment
\label{lastpage}
\end{document}